# Efficient erbium-doped thin-film lithium niobate waveguide amplifiers

Zhaoxi Chen,[1,†] Qing Xu,[2,†] Ke Zhang,[1] Wing-Han Wong,[1] De-Long Zhang,[2,3] Edwin Yue-Bun Pun,[1] and Cheng Wang[1,*]

[1]*Department of Electrical Engineering & State Key Laboratory of Terahertz and Millimeter Waves, City University of Hong Kong, Kowloon, Hong Kong, China*

[2]*Department of Opto-electronics and Information Engineering, School of Precision Instruments and Opto-electronics Engineering, Key Laboratory of Optoelectronic Information Science & Technology (Ministry of Education), Tianjin University, Tianjin, 300072, China*

[†]*These authors contributed equally to this letter*

[3]*dlzhang@tju.edu.cn*

[*]*cwang257@cityu.edu.hk*

**Abstract:** Lithium niobate on insulator (LNOI) is an emerging photonic platform with great promises for future optical communications, nonlinear optics and microwave photonics. An important integrated photonic building block, active waveguide amplifiers, however, is still missing in the LNOI platform. Here we report an efficient and compact waveguide amplifier based on erbium-doped LNOI waveguides, realized by a sequence of erbium-doped crystal growth, ion slicing and lithography-based waveguide fabrication. Using a compact 5-mm-long waveguide, we demonstrate on-chip net gain of > 5 dB for 1530-nm signal light with a relatively low pump power of 21 mW at 980 nm. The efficient LNOI waveguide amplifiers could become an important fundamental element in future lithium niobate photonic integrated circuits.

Lithium niobate on insulator (LNOI) has recently emerged as a promising material platform for compact, high-performance and low-cost photonic circuits [1–3]. Compared to conventional bulk lithium niobate (LN) optical devices based on weakly confined waveguides, LNOI devices feature smaller footprint, stronger optical confinement and significantly higher nonlinear optical and electro-optic efficiencies. To date, LNOI devices with excellent performances and a variety of functionalities have been realized, including high-speed electro-optic modulators [4–7], efficient frequency convertors [8–11], frequency comb sources [12–14], photon-pair generators [15,16], as well as on-chip spectrometers [17]. However, an important functionality in photonic integrated circuits, namely active optical gain, is typically absent in the LNOI platform, limited by the material properties of intrinsic LN.

Doping LN crystal with erbium (Er) ions is a promising solution to achieve the missing optical gain function in the LNOI platform, since $Er^{3+}$ ions could provide significant gain near the telecom wavelength range and the gain spectrum is not strongly affected by the host environment. In fact, Er-doping-based waveguide amplifiers have been realized in many popular integrated photonic platforms, including silicon nitride (SiN) [18,19], silicon (Si) [20] and other metal oxides [21]. In conventional bulk LN waveguides, Er-doped waveguide amplifiers have also been realized, often by diffusing $Er^{3+}$ ions into Ti-diffused channel or ridge waveguides [22,23]. Due to the weak optical confinement in these low-index-contrast waveguides ($\Delta n < 0.02$) and the diffusion-induced non-uniform $Er^{3+}$-ion distribution, the optical net internal gain of bulk Er-doped LN (Er:LN) waveguide amplifiers are usually < 3 dB/cm [24]. This approach, however, cannot be directly translated to the LNOI platform, since the required diffusion temperature (> 1100 °C) is too high for the LN thin films due to thermal expansion and pyroelectric issues. One approach to achieve Er:LNOI wafers is to directly grow an Er:LN crystal, followed by a standard ion-slicing process to form the Er:LNOI wafer. Using this technique, microcavity-based Er:LNOI lasers have been demonstrated very recently, showing threshold powers below 1 mW [25–27]. An alternative approach is to perform ion implantation into LNOI wafers, although the implanted wafer still needs to go through certain annealing process to activate the $Er^{3+}$ ions, and the implantation concentration reported so far is relatively low [28].

In this letter, we experimentally demonstrate an Er:LNOI waveguide amplifier with a high on-chip net gain of > 10 dB/cm at a signal wavelength of 1531.6 nm. The lithography-defined waveguides feature strong light confinement for both 980-nm pump light and 1530-nm signal light. As a result, our devices could provide high gain at relatively low pump powers (< 20 mW), and could yield efficient amplification within a large operational dynamic range (10 nW – 100 μW signal power), with measured 980-to-1530 power conversion efficiencies up to 0.2%.

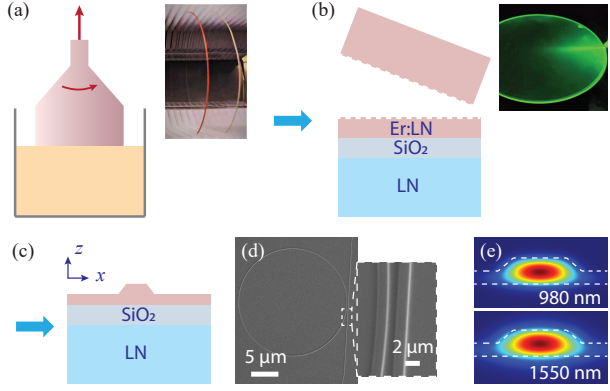

Fig. 1. (a) Czochralski process for Er:LN crystal growth. Inset shows the appearances of an Er:LN (left) and an un-doped LN (right) wafer under ambient light. (b) Ion-slicing process to create Er:LNOI wafer. Inset shows the green fluorescence generated from the Er:LN thin film when 980-nm laser light is shined from the edge (c) Cross-section schematic of the final Er:LNOI rib waveguides. (d) SEM images of the fabricated Er:LN microring resonator and bus waveguide. (e) Simulated electric field distributions ($E_x$) of the fundamental TE modes at wavelengths of 980 nm (top) and 1550 nm (bottom).

Fabrication of our Er:LNOI waveguide amplifiers starts from a bulk Z-cut LN wafer with an Er doping concentration of ~ 0.5 mol% (OST photonics). Doping of $Er^{3+}$ ions is completed during the Czochralski growth of the LN crystal, as illustrated in Fig. 1(a). The Er:LN wafer appears a color of rose red, due to the fluorescence of $Er^{3+}$ ions from ambient ultraviolet excitation [inset of Fig. 1(a)]. The bulk Er:LN wafer is then sent out for ion-slicing service (NANOLN), to transform into an Er:LNOI wafer with a top 700 nm thick Er:LN film, a 2-µm buried silica layer, and a 500-µm Z-cut un-doped LN bottom substrate, as shown in Fig.1(b). To fabricate the optical waveguides, a 250-nm-thick nichrome (NiCr) mask is first patterned through a sequence of electron-beam lithography (EBL), thermal evaporation and standard lift-off process. The mask patterns are subsequently transferred to the Er:LNOI layer by an $Ar^+$-based reactive ion etching (RIE) process. After removing the residual NiCr mask, a chemical-mechanical polishing (CMP) process is applied on the sample surface to further reduce the waveguide top and sidewall roughness [29]. The fabricated Er:LNOI waveguide has a rib-like structure, with a rib height of 320 nm, an un-etched slab thickness of 340 nm, a top width of 1.2 µm and a sidewall angle of ~ 45° [Fig. 1(c-e)]. Finally, the chip facets are carefully polished using a series of descending powder sizes to precisely control the total waveguide length (5 mm) and to produce smooth waveguide facets for efficient fiber-to-chip coupling. Figure 1(d) shows the scanning electron microscope (SEM) image of a waveguide-coupled microring resonator fabricated on the same chip as the waveguide amplifiers, revealing smooth etched surfaces. The micron-scale waveguides support strong light confinement for both 980-nm pump and 1530-nm signal light used in this work. Figure 1(e) shows the simulated field distributions ($E_x$) of the fundamental transverse electric (TE) modes at both wavelengths.

We characterize the optical amplification performance of the fabricated Er:LNOI waveguides using an end-fire coupling system shown in Fig. 2(a). Pump light from a continuous-wave (CW) 980-nm laser source (Amonics ALD-980, actual output wavelength at 974.5 nm) and signal light from a CW telecom tunable laser (Santec TSL-550) are combined using a fiber-based wavelength division multiplexer (WDM) and coupled into the Er:LNOI waveguides using a lensed fiber. In-line fiber polarization controllers are used to ensure TE polarization at the input, which shows the highest fiber-to-chip coupling efficiency in our devices. The output light is collected by a second lensed fiber and sent to an optical spectrum analyzer (OSA, YOKOGAWA AQ6370D). During our experiment processes, we observe strong up-conversion induced green photoluminescence along the Er:LNOI waveguides under test, as shown in the photograph in Fig.2(a), indicating the $Er^{3+}$ ions are well activated inside these etched structures.

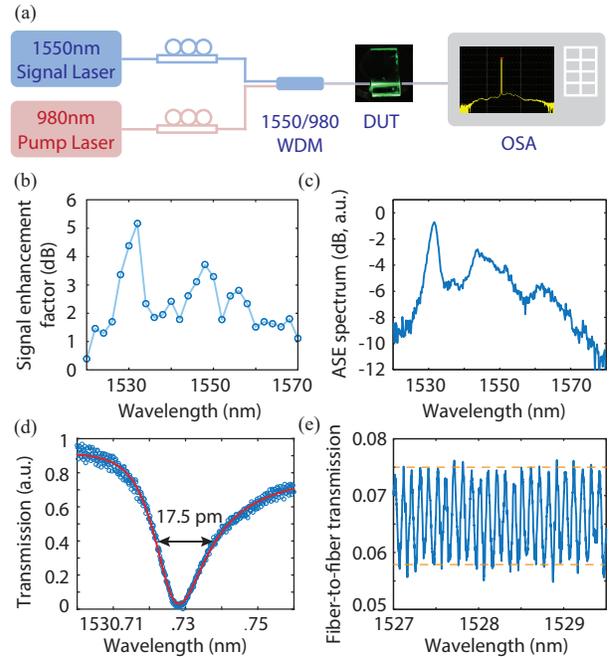

Fig. 2. (a) Schematic of the experiment setup for the gain characterization of Er:LNOI waveguide amplifiers. The photograph of the device under test (DUT) shows green fluorescence along the waveguide when pumped with 980-nm laser. (b) Signal enhancement factor as a function of signal wavelength, showing amplification from 1520 to 1570 nm. (c) Amplified spontaneous emission (ASE) spectrum of the Er:LNOI waveguide. (d) Optical transmission spectrum (blue circles) and Fano-fit curve (red line) of an Er:LNOI microring resonator. (e) Fiber-to-fiber transmission spectrum showing strong Fabry-Perot fringes.

We show that our Er:LNOI waveguides could provide signal amplification over the entire telecom C band [Fig. 2(b)]. To characterize the small-signal gain performance, we fix the input signal power at 5 nW and measure the signal enhancement factor at various wavelengths, which is defined as the output signal power ratio between pump-on (~ 20 mW) and pump-off scenarios. The powers quoted here correspond to on-chip powers, where the fiber coupling losses have been calibrated out (details to be discussed later). As Fig. 2(b) shows, the device features measurable signal amplification over the entire 1520 – 1570 nm range, with three specific enhancement peaks at the wavelengths around 1531 nm, 1546 nm and 1554 nm. These peaks agree well with the amplified spontaneous emission (ASE) spectrum of our devices [Fig. 2(c)], measured when the signal power is turned off, as well as the ASE spectrum in conventional Er-doped systems, strongly suggesting that the measured amplification phenomena in our devices are indeed the results of pump-induced $Er^{3+}$ population inversion.

To further investigate the net internal gain of our Er:LNOI waveguide amplifiers, we perform a careful calibration of the waveguide propagation loss at the signal wavelength using two corroborating measurements. We first characterize the quality ($Q$) factors of a 100-μm-radius microring resonator fabricated on the same Er:LNOI chip with the same waveguide parameters [Fig.1(d)]. Figure 2(d) shows the optical transmission spectrum of a representative resonance dip measured at 1530.73 nm. The resonance shows a slightly asymmetric line-shape due to the background Fabry-Perot interference pattern, which is induced by the reflections from the two polished chip facets. Nevertheless, the resonance spectrum can be fitted very well using a Fano function, revealing a resonance linewidth of 17.5 pm. This corresponds to an intrinsic $Q$ factor of $Q_i$ = 176,000 considering the resonator is close to critical coupling at this wavelength. We can therefore estimate the propagation loss $\alpha$ of the tested Er:LNOI waveguides using the equation

$$\alpha = \frac{2\pi N_{eff}}{\lambda_0 Q_i} \quad (1)$$

where $N_{eff}$ is the effective index of the waveguide at the target wavelength $\lambda_0$. The calculated propagation loss at 1530 nm is ~ 2 dB/cm. The second approach evaluates the waveguide loss using the contrast of the measured Fabry-Perot fringes [~ 13% as shown in Fig. 2(e)] and the numerically simulated facet reflection coefficients [30], yielding an estimated propagation loss of 1.13 dB/cm. Considering the possible difference between the simulated and actual facet reflectivity, we take the conservative but more reliable loss estimate from the first approach (*i.e.* resonator $Q$-analysis) to avoid overestimation of the net gain, which leads to an on-chip loss of 1 dB for our 5-mm device. As such, the measured fiber-to-fiber insertion loss of ~ 12 dB [Fig. 2(e)] can be attributed to a 1-dB on-chip loss and an 11-dB coupling loss, the latter of which is consistent with previous reports on similar platforms [4]. We further estimate that approximately half of the 2-dB/cm propagation loss comes from the $Er^{3+}$ ion absorption, since the typical measured $Q$ factors in our un-doped LNOI microring resonators using the same fabrication method are typically ~ 400,000. The Er absorption loss at other telecom wavelengths should in principle be lower than that at 1530 nm, but the difference is too small to be resolved in our current device. In the rest of this paper, we use a conservative 1-dB estimate of the internal loss for both 1530-nm and 1550-nm signal light.

Apart from the loss calibration at telecom wavelengths, we have also calibrated the coupling and propagation losses of our waveguides at 980 nm, by measuring and comparing the fiber-to-fiber insertion losses of the Er:LNOI waveguide and a short, un-doped LNOI waveguide with similar dimensions. The measured losses are 17 dB and 12 dB respectively, indicating a 5-dB on-chip loss and a 6-dB-per-facet coupling loss at 980 nm. The launched pump and signal powers in this paper have been calibrated to the on-chip powers using the above measurement results.

We demonstrate strong net internal gain of > 5 dB in our Er:LNOI waveguide amplifiers within a short device length and relatively low pump power. Net internal gain is calculated by subtracting the internal loss from the measured signal amplification factor. Figure 3(a) shows the measured signal spectra at 1531.6 nm at increasing pump powers (from 0 to 21 mW). A maximum net internal gain of 5.2 dB is achieved at the highest pump power of 21 mW, corresponding to a net gain per unit length of > 10 dB/cm, which is comparable with the gain achieved in other Er-doped integrated photonic systems [18–20]. When compared with typical bulk Er:LN waveguides (2 – 3 dB/cm) [22–24], our measured net internal gain is ~ 5x higher, likely due to the more efficient excitation and stimulated emission of Er ions in these strongly confined waveguides. Figure 3(c) shows the measured net internal gain as a function of pump power. At small pump powers, the optical gain increases rapidly and becomes sufficient to compensate for the internal loss at pump powers < 1 mW. The optical gain starts to see saturation behavior at pump powers > 10 mW. We have also characterized the gain performance of our Er:LNOI waveguide amplifier at the signal wavelength of 1550 nm. As shown in Figs. 3(b) and 3(d), similar signal amplification phenomena could be observed at 1550 nm, but with lower gain than 1531.6 nm. The maximum net internal gain of 1.8 dB is achieved at a pump power of 21 mW, corresponding to a net gain per unit length of 3.6 dB/cm. Further reducing the waveguide propagation loss by optimizing the fabrication process could improve the net internal gain by another 1 dB/cm.

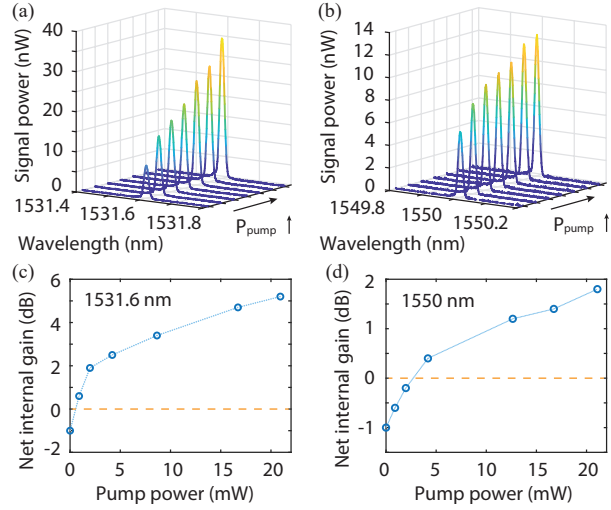

Fig. 3. (a-b) Measured signal spectra as a function of increasing pump powers at 1531.6 nm (a) and 1550 nm (b). Pump powers: 0, 1 mW, 2 mW, 4.2 mW, 8.6 mW, 16.7 mW and 21 mW. (c-d) Net internal gain as a function of pump power at 1531.6 nm (c) and 1550 nm (d).

We further characterize the operation dynamic range of our Er:LNOI waveguide amplifier by increasing the signal power until the system reaches the saturated-gain regime. With a fixed pump power of 21 mW, the net internal gain gradually drops with increasing signal power, from the small-signal value of ~ 5 dB to a final value of 0.5 dB at -10 dBm signal, as is shown in Fig. 4(a). Considering the internal loss of 1 dB, the 0.5-dB net internal gain still corresponds to a significant power amplification (~ 40% increase from a pump-off scenario), indicating a working signal dynamic range of at least 50 dB. Using these data, we could estimate the internal conversion efficiency of our Er:LNOI waveguide amplifier, defined as the percentage of the launched pump power that is transferred into signal power at the output end. Note that here the nominal gain, rather than net gain, is used to calculate the converted signal power, since the generated power for compensating the initial internal loss is also converted from 980-nm by the $Er^{3+}$ ions. As Fig. 4(b) shows, the measured conversion efficiency increases linearly with the signal power in the small-signal regime, then gradually deviates from the linear trend since the excited-state $Er^{3+}$ ions are consumed substantially by the signal light in the saturated-gain regime. At the maximum signal power of 100 μW, the internal conversion efficiency achieves a value of 0.2%, which is not yet fully saturated. We expect the saturated conversion efficiency in our current devices to be ~ 1%. The conversion efficiency as well as signal amplification performance could potentially be further improved in future work by increasing the $Er^{3+}$ doping concentration, optimizing the pump wavelength, as well as using improved waveguide designs.

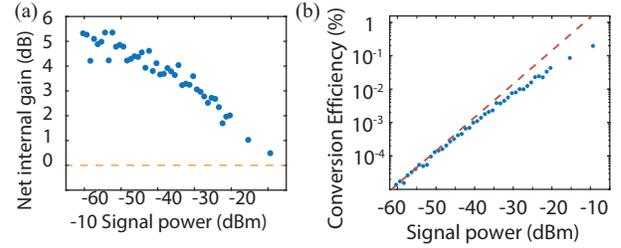

Fig. 4. (a) Dependence of net internal gain on signal power. (b) Measured internal conversion efficiency (blue dots) as a function of signal power. Red dashed line shows the fitted trend of conversion efficiency according to small-signal gain.

In conclusion, we have demonstrated an Er:LNOI waveguide amplifier with maximum net internal gain of 5.2 dB achieved in a 5-mm-long device. During the preparation of this manuscript, we noted that a similar Er:LNOI waveguide amplifier work was posted as an arXiv preprint [31], where the authors have demonstrated similar amplification results using a different waveguide fabrication approach. Our efficient and compact waveguide amplifiers could provide solutions for a long-awaiting missing piece, active gain elements, in the LNOI photonic platform, and may find applications in a variety of loss-sensitive areas such as optical communications and microwave photonics. The added gain functionality could also provide new functional degrees of freedom that could enable novel photonic systems such as parity-time symmetric systems.

**Funding.** National Natural Science Foundation of China (61922092, 61875148), Research Grants Council (Hong Kong) (CityU 11210317, CityU 21208219), City University of Hong Kong (9610402, 9610455)